\documentclass[epj]{webofc}
\usepackage[utf8]{inputenc}
\usepackage[varg]{txfonts}   % Web of Conferences font
\usepackage{booktabs}
\usepackage{xcolor}
\definecolor{darkred}{rgb}{0.4,0.0,0.0}
\definecolor{darkgreen}{rgb}{0.0,0.4,0.0}
\definecolor{darkblue}{rgb}{0.0,0.0,0.4}
\usepackage[bookmarks,linktocpage,colorlinks,
    linkcolor = darkred,
    urlcolor  = darkblue,
    citecolor = darkgreen]{hyperref}
\usepackage{subfigure}
\usepackage{url}
\usepackage{braket}
\wocname{EPJ Web of Conferences}
\woctitle{Lattice2017}
\renewcommand{\d}{\ensuremath{\mathrm{d}}}

\begin{document}
%%%%%%%%%%%%%%%%%%%%%%%%%%%%%%%%%%%%%%%%%%%%%%%%%%%%%%%%%%%%%%%%%%%%%%%%%%%%%
%
\selectlanguage{english}
%----------------------------------------------------------------------------
\title{%
Finite temperature gluon propagator in Landau gauge: 
non-zero Matsubara frequencies and spectral densities
}
%----------------------------------------------------------------------------
\author{%
\firstname{Paulo J.} \lastname{Silva}\inst{1}\fnsep\thanks{Speaker, \email{psilva@uc.pt}} \and
\firstname{Orlando} \lastname{Oliveira}\inst{1} \and
\firstname{David}  \lastname{Dudal}\inst{2,3} \and
\firstname{Martin} \lastname{Roelfs}\inst{2} 
% etc.
}
%----------------------------------------------------------------------------
\institute{%
CFisUC, Department of Physics, University of Coimbra, P--3004--516 Coimbra, Portugal
\and
KU Leuven Campus Kortrijk -- KULAK, Department of Physics, Etienne Sabbelaan 53 bus 7657, 8500 Kortrijk, Belgium
\and
Ghent University, Department of Physics and Astronomy, Krijgslaan 281-S9, 9000 Gent, Belgium
}
%----------------------------------------------------------------------------
\abstract{%
We report on the lattice computation of the Landau gauge gluon propagator at finite temperature, including the non-zero Matsubara frequencies. Moreover,  the corres\-ponding  K\"all\'en-Lehmann spectral density is computed, using a Tikhonov regularisation together with the Morozov discrepancy principle. Implications for gluon confinement are also discussed.
}
%----------------------------------------------------------------------------
\maketitle
%----------------------------------------------------------------------------
\section{Introduction and motivation}\label{intro}

The propagators of the fundamental fields of QCD, such as quarks and gluons, 
encode valuable information about non-perturbative phenomena. In particular, the gluon propagator encodes information about confinement and deconfinement. 
Following the lattice computation of the Landau gauge gluon propagator at zero temperature --- see, for example, \cite{Duarte:2016iko} and references therein --- there has also been an renewed interest in the computation of the gluon propagator at finite temperature. In fact, theoretical studies of QCD at finite temperature and density have been recently pursued due to recent heavy-ion experiments running e.g. at RHIC \cite{Trainor:2013bma} and CERN \cite{Roland:2015vgd}.
Lattice simulations of pure SU(3) Yang-Mills theory at finite temperature have found a first-order phase transition at a critical temperature $T_c\sim 270$ MeV
 \cite{Lucini:2003zr,Silva:2016onh}. Above $T_c$ the gluons become deconfined and behave as massive quasiparticles \cite{Silva:2013maa}.

In this paper we focus on the Landau gauge gluon propagator, computed from lattice simulations at finite temperature, including non-zero Matsubara frequencies. We also report on preliminary results for the spectral densities.

%----------------------------------------------------------------------------
\section{Finite temperature gluon propagator}\label{gluonprop}

The tensor structure of the Landau gauge gluon propagator, at finite temperature, includes the transverse (magnetic) $D_T$ and longitudinal (electric) $D_L$ form factors,
\begin{equation}
D^{ab}_{\mu\nu}(\hat{q})=\delta^{ab}\left(P^{T}_{\mu\nu} D_{T}(q_4,\vec{q})+P^{L}_{\mu\nu} D_{L}(q_4,\vec{q}) \right).
\label{tens-struct}
\end{equation}
Such form factors can be extracted from the calculation of $D^{aa}_{ii}(\hat{q})$ and $D^{aa}_{44}(\hat{q})$.
In this paper we consider the lattice ensembles described in \cite{Silva:2013maa}, generated in Coimbra \cite{lca} with the help of Chroma \cite{Edwards:2004sx} and PFFT \cite{pfft} libraries.

\begin{figure}[tp]
   \centering
   \subfigure[Transverse component, $T=265$ MeV.]%
             {\includegraphics[width=0.37\textwidth,clip, angle=-90]{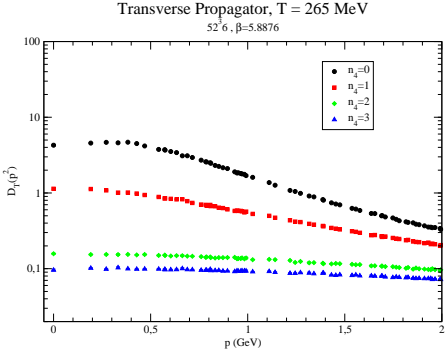}}\hfill
   \subfigure[Longitudinal component, $T=265$ MeV.]%
             {\includegraphics[width=0.37\textwidth,clip, angle=-90]{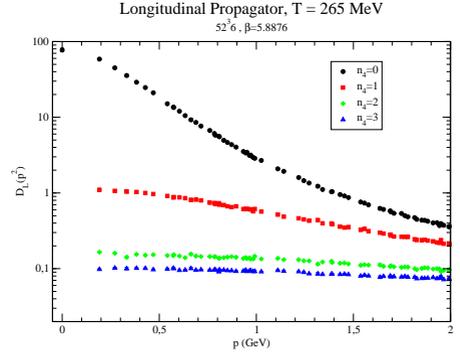}}
   \subfigure[Transverse component, $T=275$ MeV.]%
             {\includegraphics[width=0.37\textwidth,clip, angle=-90]{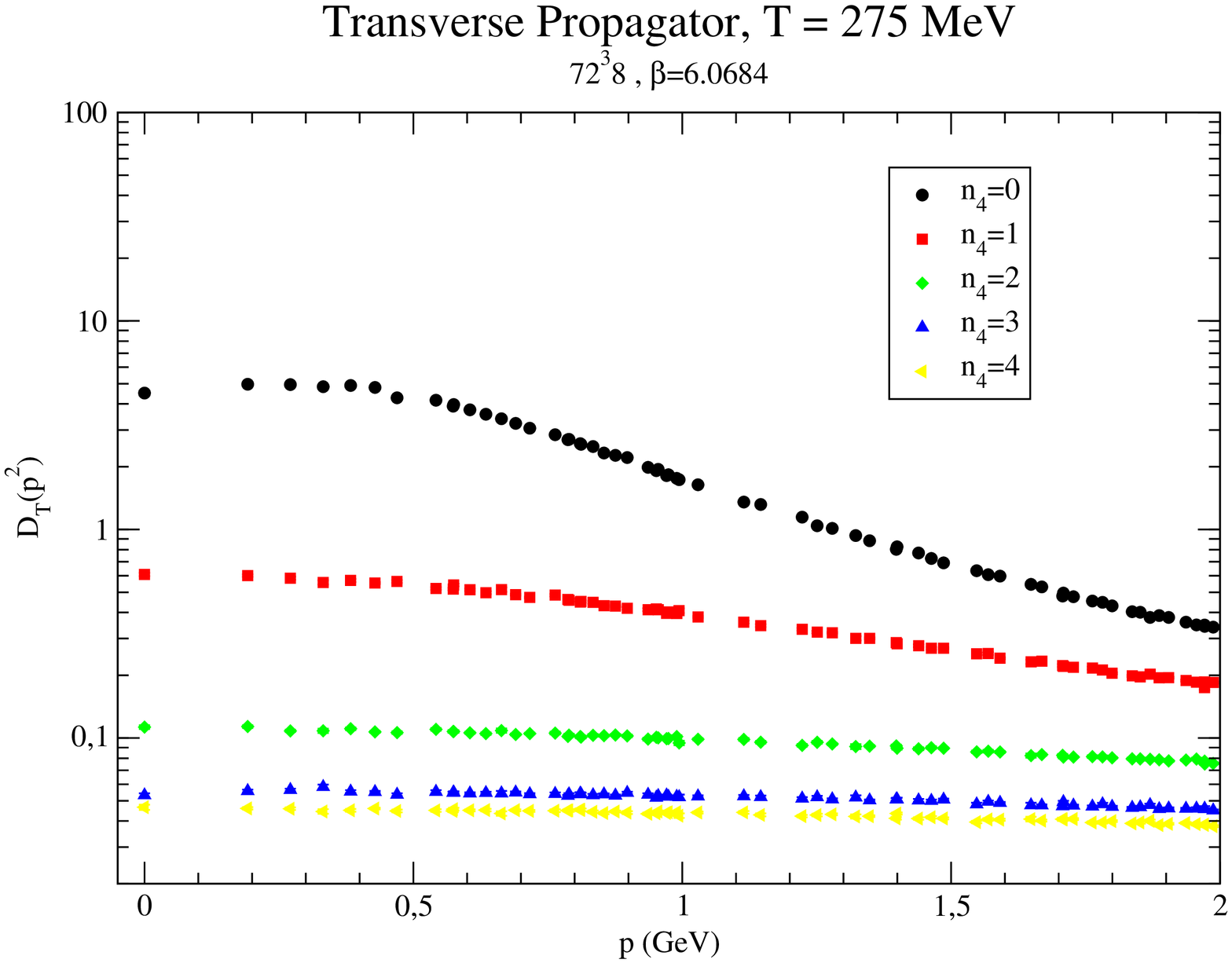}}\hfill
   \subfigure[Longitudinal component, $T=275$ MeV.]%
             {\includegraphics[width=0.37\textwidth,clip, angle=-90]{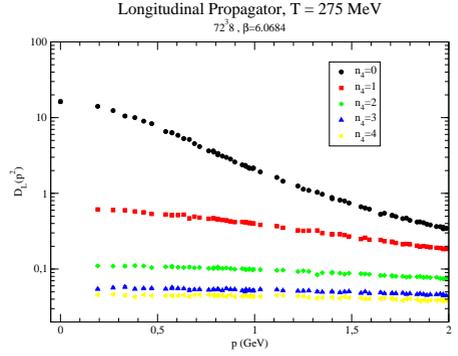}}
\hfill
   \caption{Transverse (right) and longitudinal (left) components of the gluon propagator for $T=265$ MeV (upper row) and $T=275$ MeV (lower row). }
   \label{fig:prop}
\end{figure}

In Fig. \ref{fig:prop} we see how the propagator behaves as a function of momenta. For both components and in the infrared region, the propagator decreases as we consider higher Matsubara frequencies. We therefore conclude that a larger mass scale is associated with a higher $q_4$.

\section{Spectral densities}

For some (scalar) physical degree of freedom, its Euclidean momentum-space propagator
\begin{displaymath}
\mathcal{G}(p^2)\equiv\braket{\mathcal{O}(p)\mathcal{O}(-p)}
\end{displaymath}
can be expressed in terms of the K\"{a}ll\'{e}n-Lehmann spectral representation 
\begin{displaymath}
\mathcal{G}(p^2)=\int_{0}^{\infty}\d\mu\frac{\rho(\mu)}{p^2+\mu}\,,\qquad \textrm{with }\rho(\mu)\geq0 \textrm{ for } \mu\geq 0 .
\end{displaymath}
The spectral density $\rho$ incorporates valuable information about the physical states described by the operator $\mathcal{O}$. For such physical states, $\rho$ takes positive values, and it can be computed using the maximum entropy method (MEM) \cite{Asakawa:2000tr}. 

Given that MEM requires positive spectral densities, it can not be used to extract the spectral density of unphysical degrees of freedom like gluons. Nevertheless, the method reported in \cite{Dudal:2013yva} allows both positive and negative values of the spectral density. The method relies on Tikhonov regularization combined with the Morozov discrepancy principle. 

Following \cite{Dudal:2013yva}, where we studied the gluon spectral density for $T=0$, in this paper we are interested in the finite temperature case. Unlike the case $T=0$, here one considers a single spectral function for each spatial momentum:

\begin{equation}
\mathcal{D}(q_4,\vec{q})=\int_{\mu_0}^{\infty}\d\mu\frac{\rho(\mu, \vec{q})}{q_4^2+\mu}.
\end{equation}

Lattice simulations at finite temperature rely on a compactification of the Euclidean time direction. For this reason, only a small number of Matsubara frequencies are available. However, this issue does not seem to change the main features of the extracted spectral density \cite{Silva:2017feh}. 

In this work we just report very preliminary results for the spatial momentum $\vec{p}=(1,0,0)$.

\begin{figure}[tp]
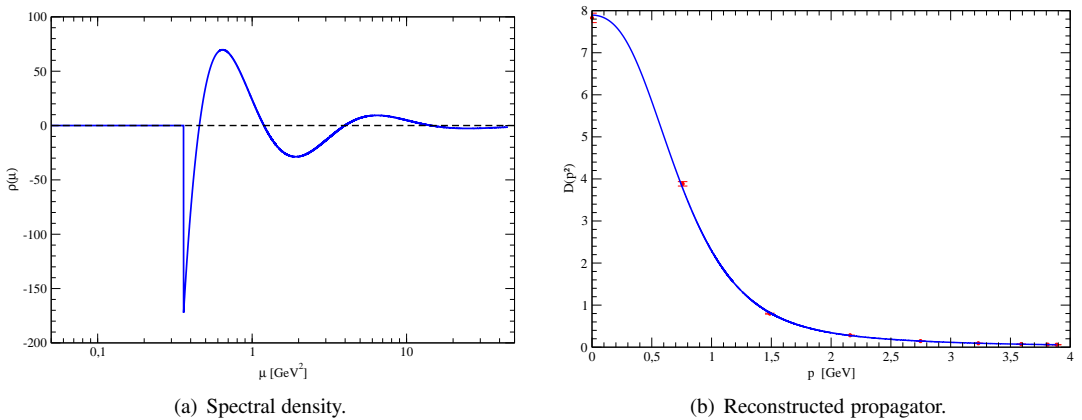

   \centering
   \subfigure[Spectral density.]%
             {\includegraphics[width=0.475\textwidth,clip]{figures/rho_trans_T121.eps}}\hfill
   \subfigure[Reconstructed propagator.]%
             {\includegraphics[width=0.475\textwidth,clip]{figures/prop_trans_T121.eps}}
   \hfill
   \caption{Results for the spectral density computed from the transverse  component of the gluon propagator at $T=121$ MeV. }
   \label{fig:rhotrans121}
\end{figure}

\begin{figure}[tp]
   \centering
   \subfigure[Spectral density.]%
             {\includegraphics[width=0.475\textwidth,clip]{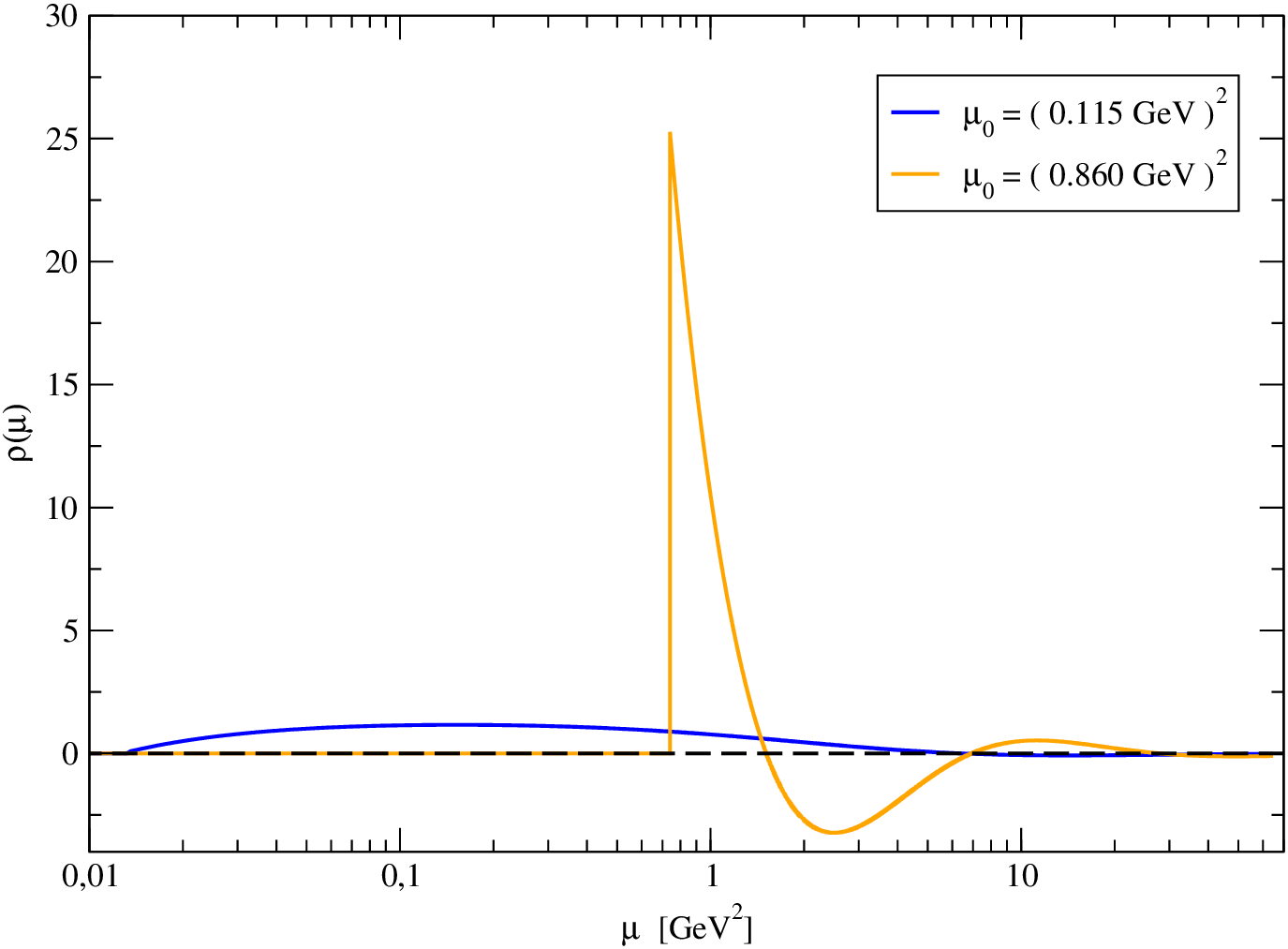}}\hfill
   \subfigure[Reconstructed propagator.]%
             {\includegraphics[width=0.475\textwidth,clip]{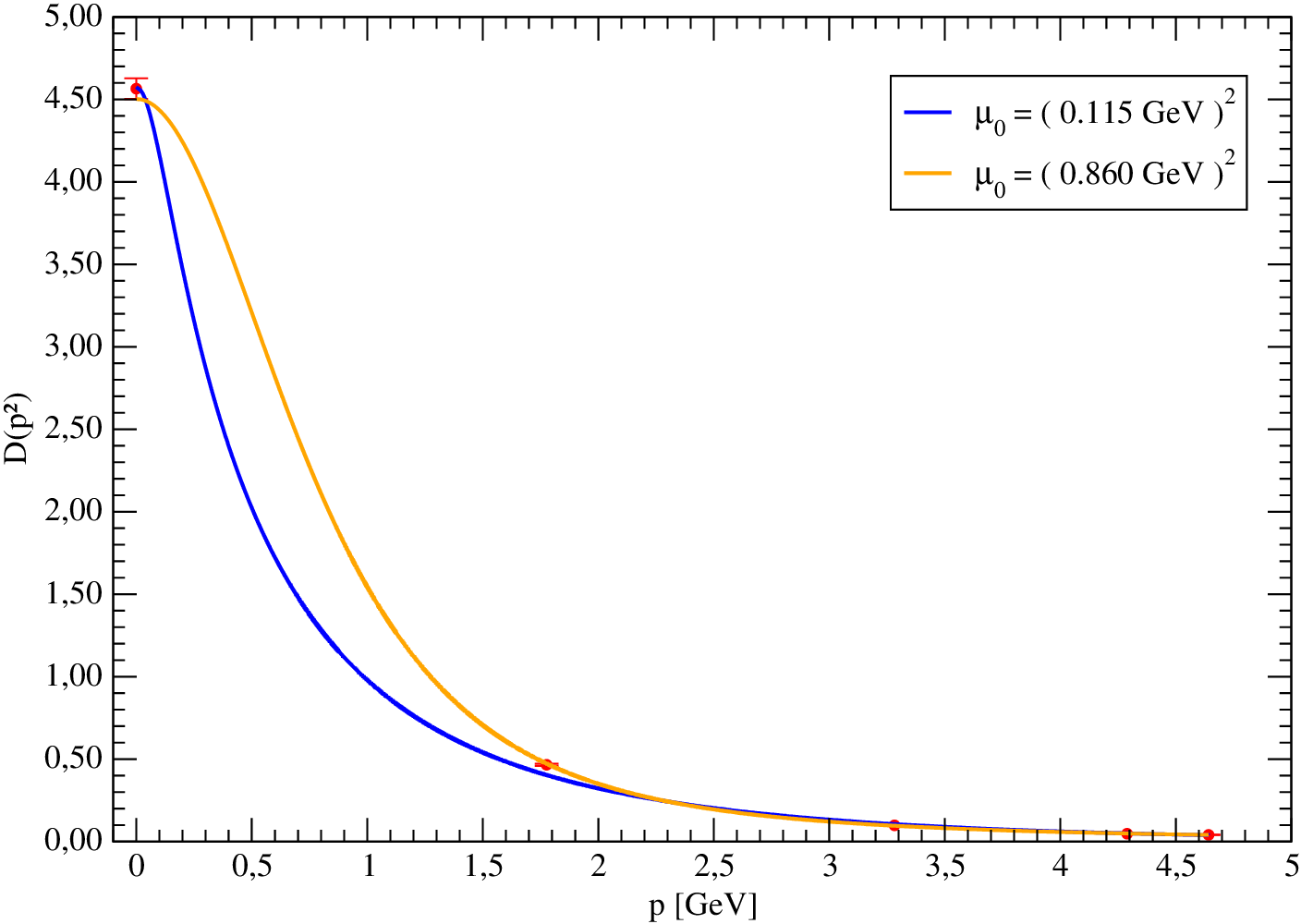}}
   \hfill
   \caption{Results for the spectral density computed from the transverse  component of the gluon propagator at $T=290$ MeV. }
   \label{fig:rhotrans290}
\end{figure}

\begin{figure}[tp]
   \centering
   \subfigure[Spectral density.]%
             {\includegraphics[width=0.475\textwidth,clip]{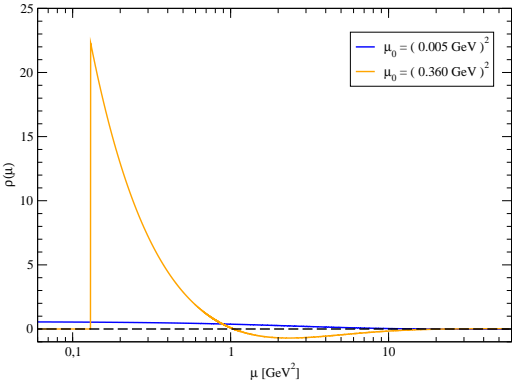}}\hfill
   \subfigure[Reconstructed propagator.]%
             {\includegraphics[width=0.475\textwidth,clip]{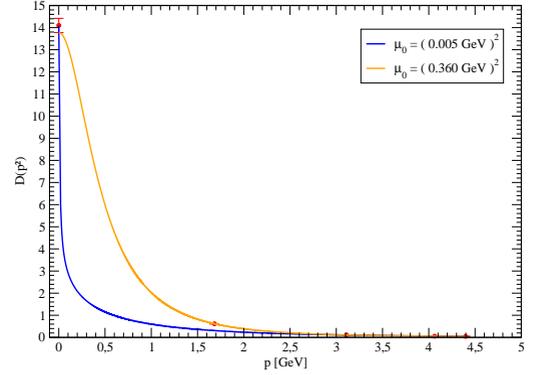}}
   \hfill
   \caption{Results for the spectral density computed from the longitudinal  component of the gluon propagator at $T=275$ MeV. }
   \label{fig:rholong275}
\end{figure}

In what concerns the transverse component, below $T_c$ the results for the spectral density share similar properties as for the $T=0$ case, as the characteristic ``sinusoidal'' behaviour --- see Fig. \ref{fig:rhotrans121}. However, 
above $T_c$ the spectral densities behave in a different way, with fewer zeroes and a big infrared cut-off $\mu_0$, as can be seen in Fig. \ref{fig:rhotrans290}. The spectral densities associated to the longitudinal component share the same pattern for almost all $T$ --- see Fig. \ref{fig:rholong275}.

\begin{figure}[tp]
   \centering
   \subfigure[Longitudinal component.]%
             {\includegraphics[width=0.475\textwidth,clip]{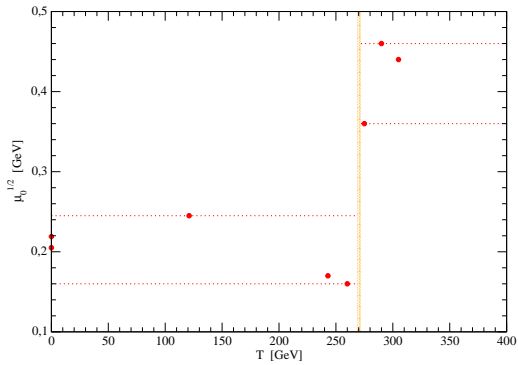}}\hfill
   \subfigure[Transverse component.]%
             {\includegraphics[width=0.475\textwidth,clip]{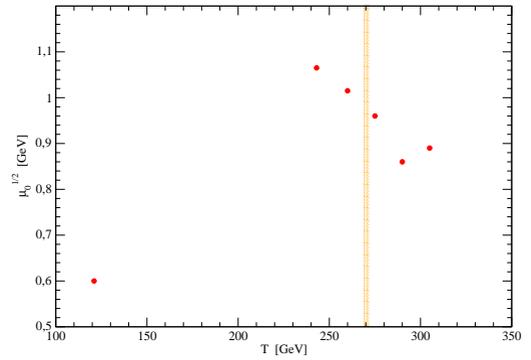}}
   \hfill
   \caption{Infrared cut-offs as functions of the temperature.}
   \label{fig:cutoffs}
\end{figure}

Fig. \ref{fig:cutoffs} shows how the infrared cut-off $\mu_0$ changes with the temperature, in particular we see that the values of $\mu_0$ change across the deconfinement phase transition.

Recently, a different method has been used to compute the spectral density \cite{Ilgenfritz:2017kkp}, with similar results.

We are working towards a full calculation of gluon spectral densities at finite temperature. 

\section*{Acknowledgments}

P. J. Silva acknowledges support by FCT under contracts SFRH/BPD/40998/2007 and SFRH/BPD/109971/2015. O. Oliveira and P. J. Silva acknowledge financial support from FCT Portugal under contract with reference UID/FIS/04564/2016. The research of M. Roelfs is funded by KU Leuven IF project C14/16/067. The computing time was provided by the Laboratory for Advanced Computing at the University of Coimbra \cite{lca}.

%\href{https://www.inspirehep.net}{inSPIRE}.  In order to cite contributions to

%\clearpage
\bibliography{lattice2017}

%%%%%%%%%%%%%%%%%%%%%%%%%%%%%%%%%%%%%%%%%%%%%%%%%%%%%%%%%%%%%%%%%%%%%%%%%%%%%
\end{document}